\documentclass[twocolumn,superscriptaddress]{revtex4-2}
\usepackage[utf8]{inputenc}
\usepackage{subfigure}
\usepackage{graphicx}
\usepackage{array}
\usepackage{xcolor}
\usepackage{amsmath}
\usepackage{amsxtra}
\usepackage{amstext}
\usepackage{amssymb}
\usepackage{amsfonts}
\usepackage{latexsym}
\usepackage{verbatim}
\usepackage{braket}
\usepackage[colorlinks=true, allcolors=blue]{hyperref}

\begin{document}

\title{Asymmetric Two-Way Gaussian Quantum Steering in Coupled Lossy Waveguides}

\author{Hafsa Zia}
\address{Department of Physics, Quaid-i-Azam University, Islamabad 45320, Pakistan}

\author{Haleema Sadia Qureshi}
\address{Department of Physics, Fatima Jinnah Women University, The Mall Rawalpindi 46000, Pakistan}

\author{Shakir Ullah}
\email{shakir@qau.edu.com}
%\thanks{These two authors contributed equally to this work.}
\address{Department of Physics, Quaid-i-Azam University, Islamabad 45320, Pakistan}
%\address{Institute of Nuclear Sciences, Hacettepe University, 06800 Ankara, Turkey}

\author{Mohamed Amazioug}
\address{Laboratory of Theoretical Physics and High Energy Physics, Department of Physics, Faculty of Sciences, Ibnou Zohr University, Agadir, Morocco}

\author{Fazal Ghafoor}
%\email{fazal\_ghafoor@comsats.edu.pk}
\address{Department of Physics, COMSATS University Islamabad, Islamabad 45550, Pakistan}

\date{\today}

\begin{abstract}
	Quantum steering is an important type of quantum correlation with potential applications in one-sided device-independent (1SDI) quantum information protocols. In this contribution, we explore the time-dependent dynamics of Gaussian quantum steering in coupled lossy optical waveguides using the covariance matrix formalism. Optical dissipation is explicitly incorporated to account for realistic propagation losses in coupled waveguides. We systematically investigate the influence of the optical loss rate, purity, nonclassicality, and squeezing of the input Gaussian states on the generation and evolution of steering. The results demonstrate that increasing nonclassicality and squeezing enhances Gaussian steering, while optical loss and reduced purity progressively suppress it. However, considerable amount of steering can persist over finite propagation intervals in the presence of optical loss. We further examine that steering is symmetric for input states with identical nonclassicality, while unequal nonclassicalities lead to directional steering asymmetry, offering a way to control the steering direction through input-state engineering. These results demonstrate that coupled optical waveguides can support controllable Gaussian quantum steering under practical loss conditions and offer a promising integrated-photonic platform for asymmetric quantum-information protocols, including 1SDI quantum key distribution (QKD) and quantum communication.
\end{abstract}
 
\maketitle

\section{Introduction}

Quantum correlations are among the fundamental features of quantum mechanics and constitute an essential resource for various advanced quantum technologies~\cite{NielsenChuang2000,AdessoRagyLee2014,HSQureshi22_TviaPC}. The development of quantum information science was initially centered on the manipulation and control of discrete quantum systems, particularly quantum bits (qubits), leading to important applications such as quantum computing~\cite{Chae2024}, quantum cryptography~\cite{Ekert1991}, and quantum teleportation~\cite{BennettBrassardCrepeauJozsaPeresWootters1993}. Continuous-variable (CV) quantum information subsequently emerged as a powerful alternative, in which information is encoded in continuous degrees of freedom of bosonic systems, such as the quadratures of the electromagnetic field~\cite{BraunsteinVanLoock2005}.

Within the CV framework, particular attention has been devoted to Gaussian states and Gaussian transformations~\cite{Weedbrook2012}, for which the quantum properties of the system can be efficiently described in terms of first and second statistical moments. This framework, commonly referred to as Gaussian quantum information processing, provides a well-established theoretical and experimental platform for implementing a broad range of quantum-information tasks~\cite{Weedbrook2012}. An important aspect of this framework is the generation, manipulation, and characterization of quantum correlations~\cite{HSQureshiBS18,JKadlec24,UllahQureshiGhafoor2019,MAmazioug18}. Among these, quantum steering~\cite{Kogias2015,SUllah19} represents a distinct form of nonclassical correlation and has attracted considerable interest due to its fundamental significance and potential applications in quantum-information protocols~\cite{Uola2020}.

Among these quantum correlations, steering has a particularly interesting historical origin. In 1935, the Einstein--Podolsky--Rosen (EPR) paradox raised the fundamental question of whether quantum mechanics provides a complete description of physical reality~\cite{einsteinPodolskyRosen1935}. Later that year, Schr\"{o}dinger introduced the concept that is now known as quantum steering~\cite{SchrodingerE35,Uola2020}. This phenomenon received renewed attention following its rigorous formulation within quantum information theory in 2007~\cite{WisemanJonesDoherty2007}. Schr\"{o}dinger originally described steering as a ``magic'' feature of quantum mechanics, whereby measurements performed by one party (Alice) can remotely steer the quantum state of another distant party (Bob), without enabling the transmission of information~\cite{Uola2020}.

In the modern formulation, steering is characterized by the impossibility of describing Bob's conditional quantum states using a local hidden state (LHS) model. Accordingly, a bipartite quantum state $\varrho_{AB}$ is said to be steerable from Alice to Bob when the correlations generated by Alice's measurements cannot be reproduced by any LHS model for Bob's subsystem. This formulation captures the intrinsically directional nature of quantum steering~\cite{KSun16,MHShah26} and distinguishes it from other forms of quantum correlations. The directional nature of quantum steering gives it a distinct operational significance in quantum information. In particular, steering provides the relevant form of nonlocal correlation for 1SDI-QKD, where the measurement apparatus of only one party is trusted~\cite{CBranciard12,NWalk16}. It also provides an operational advantage in quantum subchannel discrimination~\cite{PianiWatrous2015} and has important implications for asymmetric quantum communication and information-processing protocols~\cite{Uola2020}. These properties make the generation and preservation of steering particularly relevant to integrated photonic systems, where directional couplers and interconnected optical waveguides constitute basic building blocks for scalable quantum circuits~\cite{Politi2008,Peruzzo2010,Crespi2013}.

Optical waveguides provide compact and stable platforms for guiding and manipulating quantum states of light~\cite{Politi2008,OBrien2009}. When two waveguides are brought sufficiently close, their evanescent fields overlap, producing coherent coupling and allowing controlled exchange of optical excitations between the modes, analogous to the action of a beam splitter. Such integrated structures have been employed to realise quantum interference, quantum walks, and photonic quantum circuits~\cite{Politi2008,Bromberg2009,Peruzzo2010}. In realistic devices, however, propagation loss and coupling to environmental modes are unavoidable. These dissipative processes degrade the quantum correlations carried by the optical fields and may therefore limit the generation and preservation of steering~\cite{GardinerZoller2004,BreuerPetruccione2002}. Understanding steering under realistic loss is thus necessary for assessing its practical usefulness in integrated quantum photonic architectures.

Motivated by these considerations, we investigate the time-dependent dynamics of Gaussian quantum steering in a passive two-mode directional coupler in the presence of optical loss. Using the Gaussian-state covariance matrix formalism, we systematically examine how the loss coefficient $\kappa$, initial nonclassicality $\tau$, purity $\mu$, and input squeezing $r$ govern the generation and degradation of steering during propagation. Particular attention is given to its directional character when the two input modes possess unequal nonclassicalities. In addition to the numerical analysis, we provide an analytical treatment of the directional asymmetry. Using the exact solutions of the various moments and the local covariance matrices determinants, we demonstrate that unequal nonclassicalities of the input states generally produce unequal local covariance matrices during propagation and hence different quantum steering strengths in the two directions. This determines the imbalance of the input nonclassical resources as the origin of the asymmetric two-way steering in the proposed system. Therefore, this provides a simple mechanism for controlling steering asymmetry through input-state engineering, while simultaneously addressing the extent to which such directional quantum correlations can survive realistic propagation losses. The present study hence connects the fundamental asymmetry of Gaussian steering with the practical requirements of integrated photonic platforms for asymmetric-trust quantum-information protocols.

The paper is organized as follows. In Sec.~\ref{model_system}, we present the model system along with its Hamiltonian and dynamical equations. In Sec.~\ref{optical_loss}, we explicitly incorporate the optical loss terms. Following this, we analyze the waveguide modes and construct the covariance matrices for both input and output fields in Sec.~\ref{CMF}. In Sec.~\ref{GQSteering}, we introduce formulation of how to quantify Gaussian quantum steering. After that, we thoroughly present and discuss the graphical results in Sec.~\ref{results}, and finally the main conclusion of our results is presented in Sec.~\ref{conclusion}.

\section{Model System and Hamiltonian}\label{model_system}
\begin{figure}[t]
	\centering
	\includegraphics[width=3.2in]{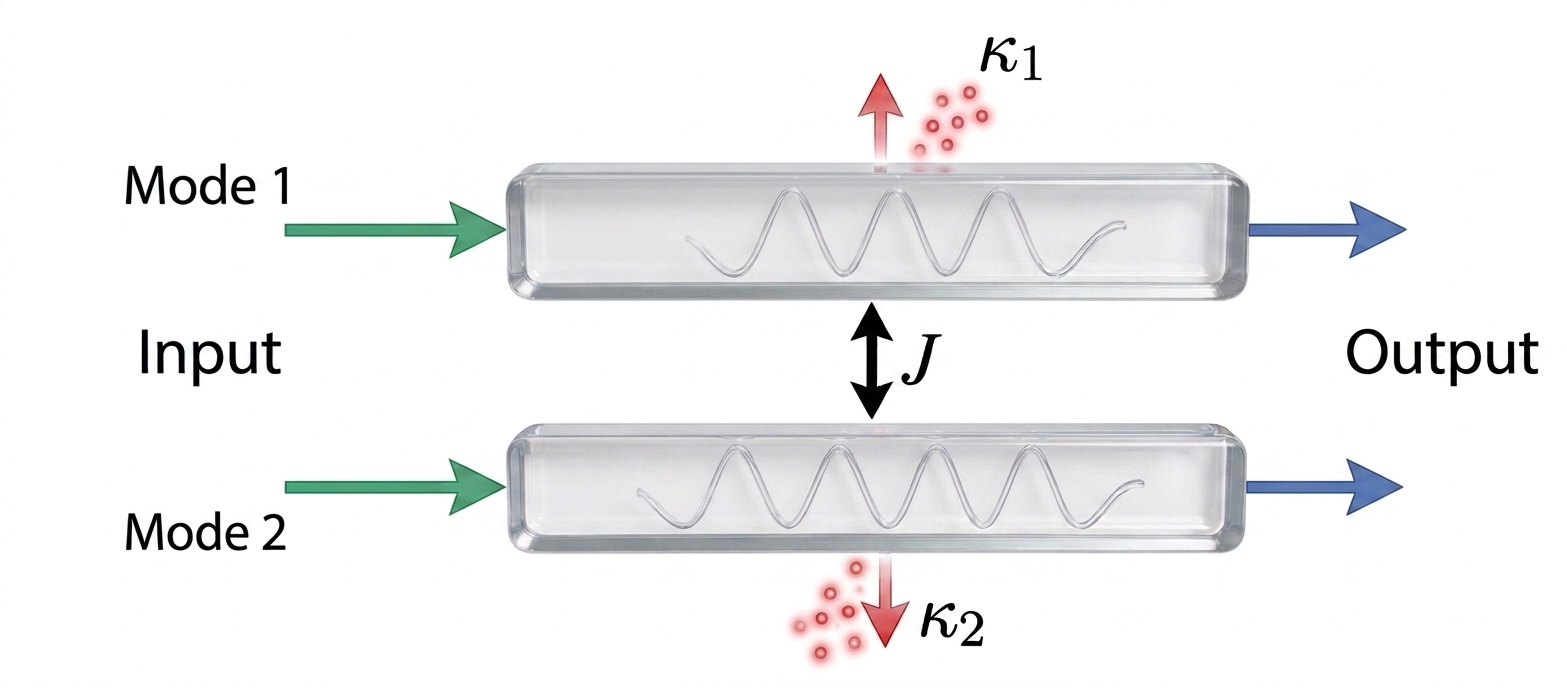}
	\caption{Schematic of coupled lossy waveguides. Two single-mode optical waveguides are placed in close proximity, enabling evanescent field coupling. The modes exchange photons coherently with coupling strength $J$ while experiencing optical loss at rate $\kappa$.}
	\label{fig:2.1}
\end{figure}

Our proposed system consists of two identical single-mode waveguides which are placed
parallel to each other in a close distance, as shown in Fig.~\ref{fig:2.1}, such that their evanescent field overlap and enables coupling between them. The quantized optical modes in the first and second waveguides are represented by the annihilation (creation) operators $a_1$ ($a_1^\dagger$) and $a_2$ ($a_2^\dagger$), respectively. Each waveguide mode acts as a bosonic harmonic oscillator. It is to be noted that
the two waveguides are assumed to be identical and at the same resonance frequency
and with the same loss rate $\kappa$, which is a realistic assumption.

The Hamiltonian for the proposed system is constructed as
\begin{equation}
	H = \hbar \omega \left(a_1^{\dagger} a_1 + a_2^{\dagger} a_2\right)
	+ \hbar J \left(a_1^{\dagger} a_2 + a_2^{\dagger} a_1\right),
\end{equation}
where $\omega$ is angular frequency of each optical mode and the parameter $J$ in the second part represents the coupling strength between the two waveguides and its value depends on the separation distance between the waveguides. Note that the bosonic operators $a_1$ and $a_2$ satisfy the commutation relations $[a_1, a_1^{\dagger}] = 1$ and $[a_2, a_2^{\dagger}] = 1$. The Hamiltonian consists of two main parts: the free energy part and the coupling part. The first part, $\hbar \omega \left(a_1^{\dagger} a_1 + a_2^{\dagger} a_2\right)$, corresponds to the free energy of the individual waveguide modes, while the second part, $\hbar J \left(a_1^{\dagger} a_2 + a_2^{\dagger} a_1\right)$, accounts for the evanescent coupling between the two waveguide modes.
%%%%%%%%%%%%%%%%%%%%%%%%%%%%%%%%%%%%%%%%%%%%%%%%%%%%%%%%%%%%%%%%%%%%%%%%%%%%%%%%%%%%%%%%%%%%%%%%%%%%%%%%%%%%%%%%%%%%%%%

\subsection{Physical Interpretation of Coupling}
The evanescent interaction between the two parallel optical waveguides is represented by the coupling term in the system Hamiltonian. When the waveguides are sufficiently close to each other, the electromagnetic field of the guided mode extends beyond the core of the waveguide and overlaps with the field of the neighboring waveguide.  This overlap allows the coherent exchange of photons between the two modes and is called evanescent overlapping~\cite{Yariv1973,SalehTeich2007}.

The strength of the coupling between the waveguides is defined by the parameter $J$ which defines the speed with which energy moves between the waveguides. The higher the value of $J$ the higher the mode overlap and higher the speed of oscillatory exchange of photons. When the two waveguides are sufficiently far apart such that the coupling parameter is zero, they are entirely independent and are considered uncoupled. Physically, such an interaction is similar to a beam splitter like interaction, where the photons coherently and simultaneously hopped between the two modes without loss in ideal case~\cite{Yariv1973,Rai2010,Politi2008}.

Without the dissipation, the coherent exchange between the two waveguides results in a periodic oscillation of photon between the two waveguides. In case of losses, such oscillations are slowly damped, although the coupling term still plays the role of creating and maintaining quantum steering. The coupling is therefore the core of creation and maintenance of quantum interactions in integrated photonic systems~\cite{Rai2010,Yariv1973}.

\subsection{Optical Loss Modeling}\label{optical_loss}
%=================================================
In realistic integrated photonic systems, optical modes inevitably experience propagation losses. To account for these losses, the coupled waveguide system is treated within the open quantum-system framework. The dynamics are described using the density matrix formalism, where optical loss is modelled through dissipative terms that account for photon leakage from the waveguide modes, resulting in a decay of the optical field and its associated quantum correlations~\cite{BreuerPetruccione2002,Rai2010}. The time evolution of the coupled waveguide system along with the losses factors is formulated in terms of the Lindblad form of the quantum Liouville equation. This approach allows inclusion of photon loss without changing the nature of the Gaussian states. This preservation is necessary for the analysis of quantum steering in the optical modes~\cite{BreuerPetruccione2002,GardinerZoller2004,Weedbrook2012,Rai2010}.

%\subsection{Quantum Liouville (Master) Equation}
The time evolution of the proposed system with loss terms is determined using the following quantum Liouville equation~\cite{BreuerPetruccione2002,GardinerZoller2004}
\begin{equation}
	\dot{\rho} = -\frac{i}{\hbar}[H, \rho] + \mathcal{L}\rho.
\end{equation}
In the above equation, the first part of the commutator generates the time evolution, where $\rho$ represents the density matrix of the system, which is responsible for the oscillatory exchange of energy and generation of quantum steering between the optical waveguide modes due to interaction parameter $J$. However, the Lindblad term ($\mathcal{L}\rho$) deals with the losses and decoherences due to interaction of system with the environment~\cite{BreuerPetruccione2002}, which is given by
\begin{equation}
	\mathcal{L}\rho = \sum_{j=1}^{2} \kappa_j \left(a_j \rho a_j^{\dagger}
	- \frac{1}{2}\left\{a_j^{\dagger} a_j, \rho\right\}\right).
	\label{eq:lindblad}
\end{equation}
Here, $\kappa_1$ and $\kappa_2$ represent the loss rates of the first and second optical modes in the waveguides~\cite{BreuerPetruccione2002}.

%\subsection{Moment Equations and Solutions}
Next, the time evolution of observable expectation values follows the form \begin{equation}
	\frac{d\langle O\rangle}{dt} = -i\langle[O,H]\rangle + \langle \mathcal{L}^\dagger(O)\rangle
	\label{eq:evolution}
\end{equation}
Note that, for CV Gaussian states, we need only second-order moments. To this end, first we calculate the time evolution for various moments, for example, for the two photon correlations and photon number and then solve these coupled differential equations with the help of initial conditions defined in Sec.~\ref{CMF}.

\subsection{Experimental Feasibility}
The physical parameter regime chosen here is in accord with experimentally accessible coupled-waveguide systems. The parameters coupling strength $J$ and optical loss rate values are experimentally reported of the orders of $J\sim10^{10}$--$10^{11}\,\mathrm{s}^{-1}$ and $\kappa\sim10^{9}$--$10^{10}\,\mathrm{s}^{-1}$, respectively, for platforms such as lithium niobate (LiNbO$_3$), AlGaAs, and silica waveguides~\cite{Iwanow2004,Peschel2002,Mogensen2004}. These experimental values correspond approximately to normalized optical loss rates $\kappa/J$ in the range $0.02$--$0.14$. Therefore, the considered values of $\kappa/J$ in our analysis show experimentally relevant low-to-moderate loss regimes, while allowing the influence of optical dissipation on Gaussian quantum steering to be examined systematically. Note that, single-mode Gaussian input states can be experimentally generated and controlled using CV optical techniques, see Sec.~\ref{GQSteering} for detail.

\section{Covariance Matrix Formalism}\label{CMF}
A single-mode Gaussian state (SMGS) is fully characterized by the parameters of its covariance matrix. The diagonal element, lets say $n_j$, of the covariance matrix quantifies the local quadrature variance of the $j$-th mode and is directly related to the mean photon number according to
\begin{equation}
	\langle a_j^{\dagger} a_j \rangle_0 = n_j - \frac{1}{2},
	\label{eq:3.20}
\end{equation}
where the term $\frac{1}{2}$ accounts for the unavoidable vacuum or zero point fluctuations of the bosonic field~\cite{Serafini2017}. Within the nonclassicality--purity parametrization of SMGS, the same quantity $n_j$ can be expressed in terms of the nonclassicality $\tau_j$ and the purity $\mu_j$ as~\cite{AdessoRagyLee2014}
\begin{equation}
	n_j = \frac{1}{2} + \frac{\tau_j^2 + \dfrac{1}{(2\mu_j)^2} - \dfrac{1}{4}}{1 - 2\tau_j},
	\label{eq:3.21}
\end{equation}
which ensures that the corresponding covariance matrix satisfies the Heisenberg uncertainty principle~\cite{Serafini2017}. The off-diagonal covariance parameter, lets say $m_j$, of the covariance matrix is defined through the anomalous second-order moment
\begin{equation}
	\langle a_j a_j \rangle_0 = -m_j,
	\label{eq:3.22}
\end{equation}
where $m_j = |m_j| e^{i\phi_j}$, in general, is a complex quantity. Within the nonclassicality--purity parametrization of SMGS, the element $m_j$ can be expressed as
\begin{equation}
	|m_j| = \frac{\tau_j - \tau_j^2 + \dfrac{1}{(2\mu_j)^2} - \dfrac{1}{4}}{1 - 2\tau_j}.
	\label{eq:3.23}
\end{equation}
These two parameters, $n_j$ and $m_j$, together fully specify the SMGS at the input, with $n_j$ determining the phase-insensitive fluctuations and $m_j$ encoding the phase-sensitive correlations of the field ~\cite{Li2006}.

In general, the covariance matrix for a two-mode state of an evolved system is given by
\begin{equation}
	\sigma(t)=
	\begin{pmatrix}
		M(t) & L(t)\\
		L^\dagger(t) & N(t)
	\end{pmatrix}.
	\label{eq:CM_block}
\end{equation}
In the above matrix, the $2 \times 2$ submatrices
\begin{equation}
	M(t)=\begin{pmatrix}
		\langle a_1^{\dagger} a_1 \rangle_t + \frac{1}{2} & -\langle a_1 a_1 \rangle_t \\
		-\langle a_1 a_1 \rangle^\ast_t & \langle a_1^{\dagger} a_1 \rangle_t + \frac{1}{2}
	\end{pmatrix},
\end{equation}
and
\begin{equation}	
	N(t)=\begin{pmatrix}
		\langle a_2^{\dagger} a_2 \rangle_t + \frac{1}{2} & -\langle a_2 a_2 \rangle_t \\
		-\langle a_2 a_2 \rangle^\ast_t & \langle a_2^{\dagger} a_2 \rangle_t + \frac{1}{2}
	\end{pmatrix},
	\label{detMN}
\end{equation}
represent the local variations in the first and second modes, while the $2 \times 2$ submatrix
\begin{equation}
	L(t)=\begin{pmatrix}
		\langle a_1 a_2^{\dagger} \rangle_t & -\langle a_1 a_2 \rangle_t \\
		-\langle a_1 a_2 \rangle_t^\ast & \langle a_1 a_2^{\dagger} \rangle_t^\ast
	\end{pmatrix},
\end{equation}
represents the correlation between the modes. Here, $t$ in the subscript represents expectation values after certain time. Next, we obtain the matrix elements of the above submatrices, which are in fact the expectation values of the two photon correlations and photon number. First, we apply Eq.~(\ref{eq:evolution}) to calculate the time evolution for various moments and then solve the obtained coupled differential equations using the initial conditions given in Eqs.~(\ref{eq:3.20}), (\ref{eq:3.22}), and  $\langle a_1^{\dagger} a_2 \rangle_0  = 0$, and $\langle a_1 a_2 \rangle_0 = 0$. The last two initial conditions show that the input modes are independently prepared and there is no inter-mode correlations. For convenience, we assume equal loss rates for identical waveguides, that is $\kappa_1 = \kappa_2 = \kappa$. The solutions are as follows
\begin{align}
	n_1(t) &= \frac{e^{-\kappa t}}{2}\Big[(n_1+n_2-1) + (n_1-n_2)\cos(2Jt)\Big], \label{eq:n1} \\
	n_2(t) &= \frac{e^{-\kappa t}}{2}\Big[(n_1+n_2-1) - (n_1-n_2)\cos(2Jt)\Big], \label{eq:n2} \\
	c(t) &= \frac{i\,e^{-\kappa t}}{2}\,(n_2-n_1)\sin(2Jt),\\
	m_1(t) &= \frac{e^{-\kappa t}}{2}\Big[(m_1+m_2)\cos(2Jt) + (m_1-m_2)\Big], \label{eq:m1} \\
	m_2(t) &= \frac{e^{-\kappa t}}{2}\Big[(m_1+m_2)\cos(2Jt) + (m_2-m_1)\Big], \label{eq:m2} \\
	s(t) &= -\frac{i\,e^{-\kappa t}}{2}\,(m_1+m_2)\sin(2Jt),
	\label{eq:moment_solutions}
\end{align}
where $n_1(t) = \langle a_1^{\dagger} a_1 \rangle_t$, $n_2(t) = \langle a_2^{\dagger} a_2 \rangle_t$, $m_1(t) = \langle a_1 a_1 \rangle_t$, $m_2(t) = \langle a_2 a_2 \rangle_t$, $c(t) = \langle a_1^{\dagger} a_2 \rangle_t$, and $s(t) = \langle a_1 a_2 \rangle_t$.
Here, $n_j$ represents photon number, $m_j$ represents anomalous moments, $c(t)$ and $s(t)$ represent inter-mode correlations, and all contain the exponential decay factor $e^{-\kappa t}$ due to loss.

The covariance matrix in Eq.~\eqref{eq:CM_block} fully characterizes the Gaussian state, and provides the basis for identifying different kinds of quantum correlations in various systems~\cite{HSQureshi21_ENinCEL,NChabarSUllah26,Harraf2025,NChabarHSQureshi26}.
This block-structured covariance matrix serves as the starting point for evaluating quantum steering in our proposed system.

\section{Gaussian Quantum Steering}\label{GQSteering}
Quantum steering characterizes the ability of one party to remotely influence or steer the quantum state of another specially separated party through local measurements~\cite{WisemanJonesDoherty2007,Uola2020,HSQureshi20}. In the present system, we consider the two-mode Gaussian state generated at the output of the coupled lossy waveguides, whose covariance matrix $\sigma(t)$ together with its block matrices, was introduced in Eq.~(\ref{eq:CM_block}). Here, the sub-covariance matrices $M(t)$ and $N(t)$ are the subcovariance matrices describe the local properties of the two subsystems, while $L(t)$ accounts for the correlations between them.

To investigate the quantum steering from one mode to another, it is convenient to group the corresponding quadrature operators $\hat{x}_{1,2}$ and $\hat{p}_{1,2}$ into a single phase-space vector $\hat{Y} = \left(\hat{x}_1,\ \hat{p}_1,\ \hat{x}_2,\ \hat{p}_2\right)^T$. These operators satisfy the canonical commutation relations. $[\hat{Y}_i, \hat{Y}_j] = i\Omega_{ij}$, where $\hat{Y}_i$ and $\hat{Y}_j$ are the $i^{th}$ and $j^{th}$ components of the quadrature operator vector defined in phase
space and $\Omega_{ij}$ denotes the $(i^{th},j^{th})$ element of the symplectic
matrix 
\begin{equation}
	\Omega = \bigoplus_{k=1}^{2}
	\begin{pmatrix} 0 & 1 \\ -1 & 0 \end{pmatrix}.
\end{equation}
Note that, the covariance matrix $\sigma(t)$ describing a physically admissible quantum state must satisfy the Robertson–Schrodinger uncertainty relation
\begin{equation}
	\sigma(t) + i\Omega \geq 0,
\end{equation}
where the inequality indicates that the matrix on the left-hand side is positive semidefinite. This condition guarantees the physicality of the underlying quantum state. Remarkably, a bipartite Gaussian state is non-steerable with Gaussian measurements performed on mode 1 cannot steer mode 2 if and only if a specific matrix inequality involving the covariance matrix $\sigma(t)$ is satisfied~\cite{WisemanJonesDoherty2007}
\begin{equation}
	\boldsymbol{\sigma}(t) + i \left( 0_1 \oplus \Omega_2 \right) \geq 0,
	\label{eq:covariance-criterion}
\end{equation}

For Gaussian measurements performed on mode 1, the steering quantifier can be expressed in terms of the Schur complement of mode 1, $R_N = N - L^{T}M^{-1}L$, whose symplectic eigenvalue satisfies $\nu_N^2 = \det R_N$.
Consequently, the Gaussian quantum steering measure from mode 1 to mode 2 is given by
\begin{equation}
	S_{1\to2}
	=
	\max\left\{0,-\ln\left(\nu_N^2\right)\right\}.
\end{equation}
In general, an equivalent expression for the degree of Gaussian steering from mode~1 to mode~2, with a
covariance matrix $\sigma(t)$, is quantified by~\cite{Kogias2015,Simon2000,Adesso2007}
\begin{equation}
	S_{1\to2}
	=
	\max\!\left[
	0,\;
	\frac{1}{2}\ln
	\frac{\det M(t)}{\det \sigma(t)}
	\right].
	\label{eq:St(1-2)}
\end{equation}
Likewise, the degree of Gaussian steering from mode~2 to mode~1 is quantified by~\cite{Kogias2015,Simon2000,Adesso2007}
\begin{equation}
	S_{2\to1}
	=
	\max \left[
	0,	\frac{1}{2}\ln
	\frac{\det N(t)}{\det \sigma(t)}
	\right].
	\label{eq:St(2-1)}
\end{equation}
\subsection{Analytical Treatment of Quantum Steering Asymmetry}\label{sec:AsySt}

The origin of the directional asymmetry can be easily derived directly from the Gaussian steering measures given in Eqs.~(\ref{eq:St(1-2)}) and (\ref{eq:St(2-1)}). For nonzero steering in both the directions, their difference is written as
\begin{equation}
	\Delta S(t)
	\equiv
	S^{1\rightarrow2}(t)
	-
	S^{2\rightarrow1}(t)
	=
	\frac{1}{2}
	\ln\left[
	\frac{\det M(t)}{\det N(t)}
	\right].
	\label{eq:steering_difference}
\end{equation}
From the above expression, symmetric steering requires $\det M(t)=\det N(t)$, while unequal local covariance determinants certainly give rise to distinct steering strengths in the two directions. Hence, the difference between the local covariance matrices, defined in Eq.~(\ref{detMN}), is expressed as
\begin{align}
	\det M(t)-\det N(t)
	={}&
	\big[n_1(t)-n_2(t)\big]
	\big[n_1(t)+n_2(t)+1\big]
	\nonumber\\
	&-
	\left[
	|m_1(t)|^2-|m_2(t)|^2
	\right].
	\label{eq:det_difference_initial}
\end{align}
Substituting the exact solutions for $n_1(t)$, $n_2(t)$, $m_1(t)$, and $m_2(t)$ given in Eqs.~(\ref{eq:n1}), (\ref{eq:n2}), (\ref{eq:m1}), and (\ref{eq:m2}), respectively, we obtain
\begin{align}
	\det M(t)-\det N(t)
	=&
	e^{-\kappa t}\cos(2Jt)
	\Bigg\{
	(n_1-n_2)
	\nonumber\\
	&
	\times \left[
	1+e^{-\kappa t}(n_1+n_2-1)
	\right]
	\nonumber\\
	&
	-
	e^{-\kappa t}
	\left(
	|m_1|^2-|m_2|^2
	\right)
	\Bigg\}.
	\label{eq:det_difference_final}
\end{align}
Eq.~(\ref{eq:det_difference_final}) provides a direct analytical explanation of the asymmetric nature of steering in terms of the two input Gaussian state parameters. On one hand, for pure input states, when the two input modes have equal nonclassicalities, that is $\tau_1=\tau_2$, Eqs.~(\ref{eq:3.21}) and (\ref{eq:3.23}) give
\begin{equation}
	n_1=n_2,
	\qquad
	|m_1|=|m_2|.
\end{equation}
Hence from Eq.~(\ref{eq:det_difference_final}), we obtain
\begin{equation}
	\det M(t)=\det N(t),
\end{equation}
and accordingly
\begin{equation}
	S^{1\rightarrow2}(t)
	=
	S^{2\rightarrow1}(t),
\end{equation}
which corresponds to symmetric two-way steering.

On the other hand, when the two input modes have unequal nonclassicalities, that is $\tau_1\neq\tau_2$, while keeping the input purities equal, Eqs.~(\ref{eq:3.21}) and (\ref{eq:3.23}) generally give
\begin{equation}
	n_1\neq n_2,
	\qquad
	|m_1|\neq|m_2|.
\end{equation}
Therefore, the two terms on the right-hand side of Eq.~(\ref{eq:det_difference_final}) generally produce
\begin{equation}
	\det M(t)\neq\det N(t),
\end{equation}
and consequently
\begin{equation}
	S^{1\rightarrow2}(t)
	\neq
	S^{2\rightarrow1}(t),
	\label{eq:asymmetric_steering}
\end{equation}
for generic propagation times. This shows that the directional asymmetry of quantum steering observed in the present system originates from the unequal nonclassical resources of the two input Gaussian modes. Since both waveguides experience the same optical loss rate, $\kappa_1=\kappa_2=\kappa$, the asymmetry in steering is an input-state effect rather than a consequence of unequal dissipation. Moreover, swapping $\tau_1$ and $\tau_2$ inverts the input imbalance and accordingly reverses the directional nature of steering, which is fully consistent with the numerical results illustrated in Figs.~\ref{fig:4.8}(a) and \ref{fig:4.8}(b).

\subsection{Definition of Nonclassicality and Purity}
It is worth mentioning that we characterize the input Gaussian states using two key parameters, for instance, the nonclassicality $\tau_i$ and the purity $\mu_i$, with $i=1,2$~\cite{SUllah19_EninRDQBL}. The nonclassicality parameter is defined as $\tau_i=\max \{0,\frac{1}{2}-\lambda_{\text{min}}^i\}$, where $\lambda_{\text{min}}^i$ is the minimum eigenvalue of covariance matrix, for example $V_{\text{in}}^i$, of input SMGS. $\tau_i$ ranges from $0$ to $\frac{1}{2}$ and is related to the squeezing properties of the input state and controls the degree of deviation from classical behavior. Here, $\tau=\frac{1}{2}$ corresponds to a fully nonclassical state. However, the purity $\mu \in [0,1]$ quantifies the degree of mixedness of a quantum state, defined as $\mu =1/2\sqrt{\det V_{\text{in}}^i}$, where $\mu=1$ corresponds to a pure state and $\mu<1$ indicates a mixed state. These two meaningful parameters are independently controllable in our setup. All values of these $\tau_i$, and $ \mu$ used in the simulations satisfy the physical realizability constraints for two-mode Gaussian states~\cite{HSQureshi23}. 

From an experimental perspective, the single-mode Gaussian input states considered here can be prepared using existing CV optical techniques. Their quadrature properties can be manipulated via optical squeezing, displacement operations, and phase-space rotations, whereas various degrees of purity can be achieved by incorporating controlled optical loss or Gaussian noise. Gaussian states have been experimentally created and controlled using optical parametric processes, with their squeezing and other system parameters being tunable in the experimental configuration~\cite{Wenger2004,Dauria2009}. The generated states can be characterized by balanced homodyne detection process, which enables reconstruction of the relevant quadrature moments and covariance matrix and, accordingly, evaluation of quantities for example squeezing, purity, and nonclassicality~\cite{Dauria2009,Buono2010}. Thus, the input Gaussian states chose in our analysis can be associated with experimentally preparable, controllable, and measurable optical states.
%%%%%%%%%%%%%%%%%%%%%%%%%%%%%%%%%%%%%%%%%%%%%%%%%%%%%%%%%%%%%%%%%%%%%%%%%%%%%%%%
\begin{figure}[t]
	\centering
	\includegraphics[width=2.8in]{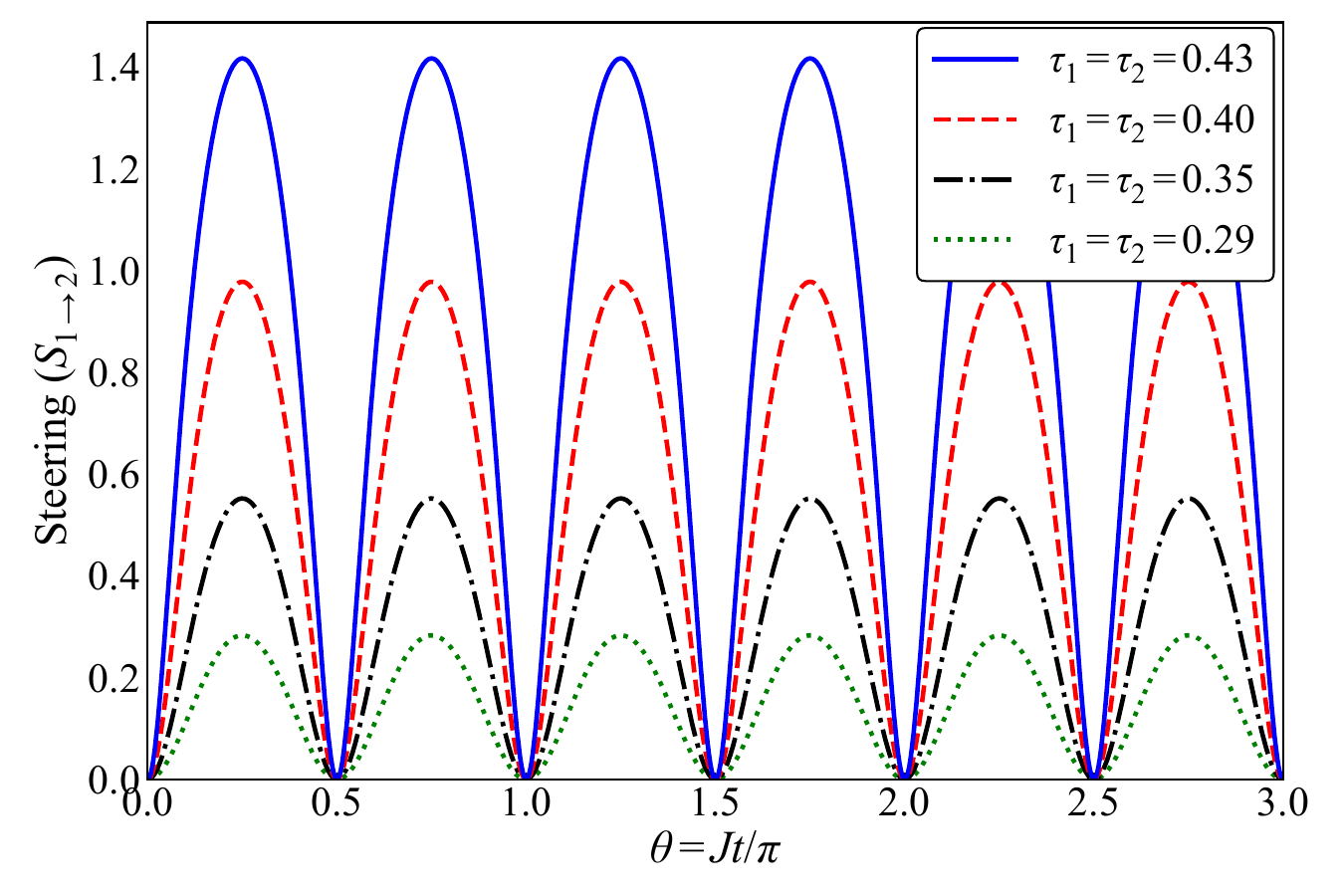}
	\caption{Effect of nonclassicality on Gaussian quantum steering is plotted against dimensionless interaction time, for fixed parameter values $\kappa_1 = \kappa_2 = \kappa = 0.0$, $\mu_1 = \mu_2 = \mu = 1$, $\phi_1 = 0$ and $\phi_2 = \pi/2$. In each case, dotted-green, dot-dashed black, dashed-red, and solid blue lines represent results $\tau_1 = \tau_2 = \tau = 0.29, 0.35, 0.40$, and $0.43$, respectively.}
	\label{fig:4.2}
\end{figure}

\section{Results and Analysis}\label{results}

In this section, we investigate the generation of Gaussian quantum steering during the propagation of the optical fields through the coupled lossy waveguides. The time evolution of steering are controlled by the coherent coupling between the waveguide, the nonclassicality and purity of the input states, and the optical loss rates. In the following, we examine how the nonclassicality $\tau$, optical loss $\kappa$, squeezing, and purity $\mu$ affect the magnitude and time development of steering. In addition, we explore the asymmetric nature of quantum steering when the two input Gaussian modes have different nonclassicalities.  
%%%%%%%%%%%%%%%%%%%%%%%%%%%%%%%%%%%%%%%%%%%%%%%%%%%%%%%%%%%%%%%%%%%%%%%%%%%%%%%%%%%%%%%%%%%%%%%%%
\begin{figure}[t]
	\centering
	\includegraphics[width=2.8in]{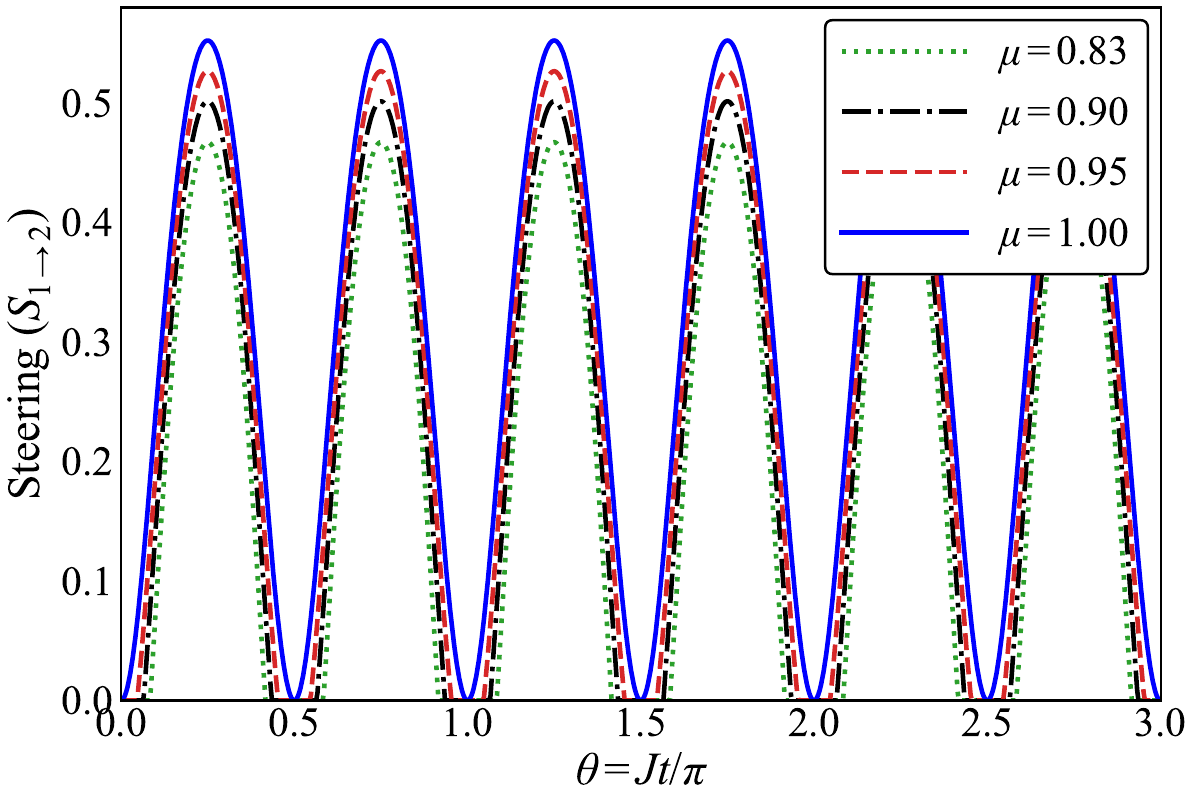}
	\caption{Effect of purity on Gaussian quantum steering is plotted against dimensionless interaction time, for fixed parameter values $\kappa_1 = \kappa_2 = \kappa = 0.0$, $\tau_1 = \tau_2 = \tau = 0.35$, $\phi_1 = 0$ and $\phi_2 = \pi/2$. In each case, dotted-green, dot-dashed black, dashed-red, and solid blue lines represent $\mu_1 = \mu_2 = \mu = 0.83, 0.90, 0.95$, and $1.0$, respectively.}
	\label{fig:4.6}
\end{figure} 

The effect of input state's nonclassicality on the Gaussian quantum steering is shown in Fig.~\ref{fig:4.2}, where the initial two modes are prepared equally nonclassical, that is $\tau_1=\tau_2=\tau$, while the remaining parameters are kept fixed. Here, we consider $\kappa=0$, thus the evolution is fully controlled by the coherent coupling between the two waveguides. Therefore, the steering exhibits an oscillatory and periodic profile with respect to the dimensionless interaction time. This oscillatory behavior emerges due to the continuous exchange and redistribution of the quantum correlations between the two coupled optical modes. From the figure, a clear enhancement of quantum steering is seen, where the maximum value increases monotonically with the increase of input modes nonclassicality. However, the positions of the maxima and minima remain the same as the period of oscillation is mainly determined by the coherent coupling strength $J$, rather than by input nonclassicality $\tau$. The parameter $\tau$ thus primarily controls the strength of quantum resource responsible for the generation of steering, while the coupling parameter $J$ controls its periodic exchange between the two modes. These findings demonstrate that increasing the nonclassical character of the input states provides an effective means of enhancing Gaussian quantum steering in coupled waveguides.

In Fig.~\ref{fig:4.6}, the dependence of purity $\mu$ of the input states on Gaussian quantum steering is illustrated, with the nonclassicality fixed at $\tau=0.35$ and in the absence of optical dissipation. The maximum amount of steering is achieved when the inputs states are pure, that is $\mu=1$, whereas the amount is found to be reduced with the decrease purity. Note that this reduction is due to the increased statistical noise associated with mixed input states, which weakens the quadrature correlations responsible for the generation of Gaussian steering between the modes. Nevertheless, the positions of the maxima and minima of the steering profile remain the same, indicating that purity primarily influence the steering strength rather than the characteristic time scale of its coherent evolution. In addition to this, a considerable strength of steering sustains even for moderately mixed input states, which is promising for realistic optical systems where perfectly pure Gaussian states are difficult to generate.

\begin{figure}[t]
	\centering
	\includegraphics[width=2.8in]{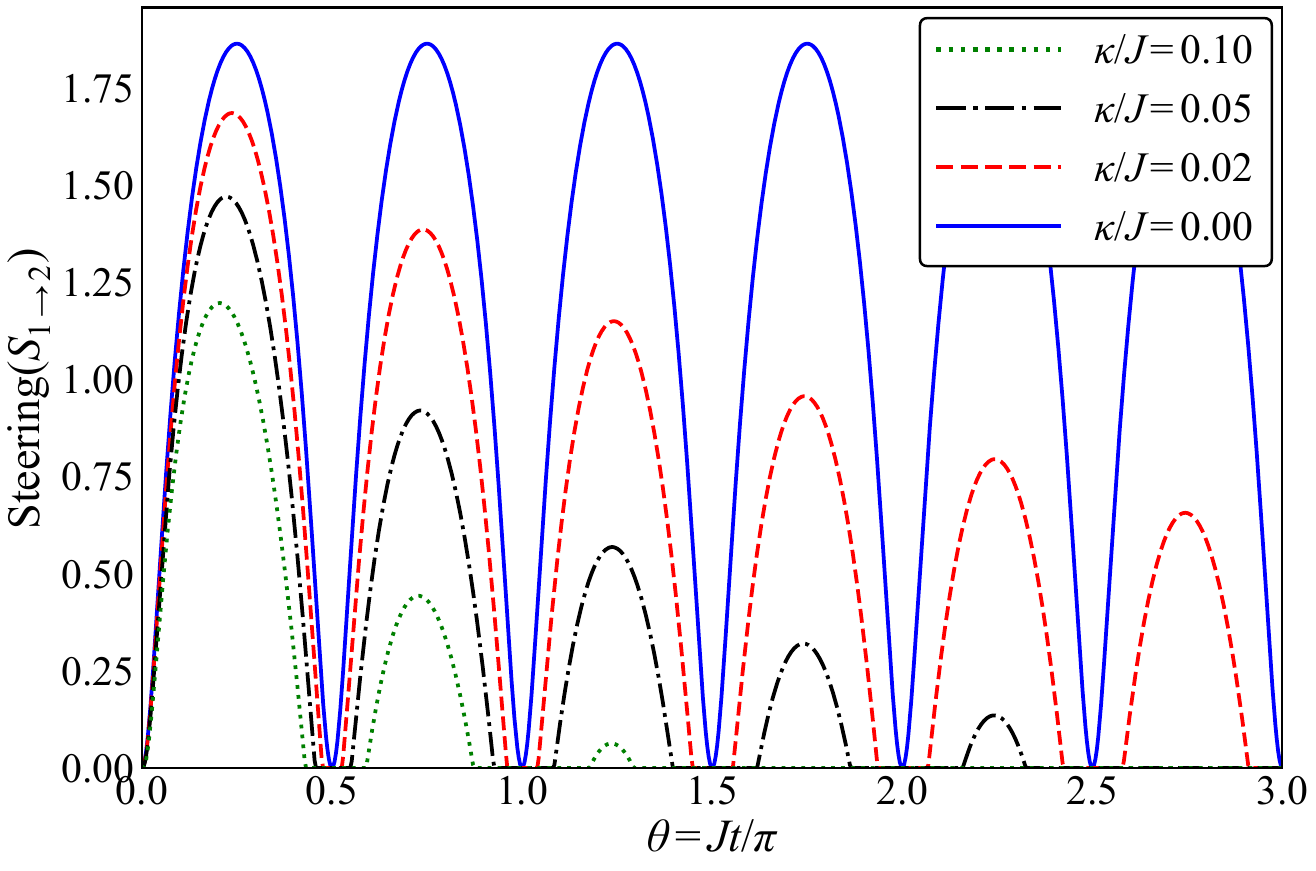}
	\caption{Effect of optical loss is plotted against dimensionless interaction time, for fixed parameter values $\tau_1 = \tau_2 = \tau = 0.45$, $\phi_1 = 0$ and $\phi_2 = \pi/2$, and $\mu_1 = \mu_2 = \mu = 1$. In each case, dotted-green, dot-dashed black, dashed-red, and solid blue lines represent results for $\kappa/J = 0.10, 0.05, 0.02$, and $0.00$, respectively.}
	\label{fig:4.3}
\end{figure}

The influence of optical loss rate $\kappa/J$ on Gaussian quantum steering is shown in Fig.~\ref{fig:4.3}, whereas the remaining parameters, such as nonclassicality, purity, and relative phases of the input states are kept fixed. For lossless case when $\kappa/J=0$, represented by the solid blue curve, the steering exhibits persistent oscillations without a decay the profile. This shows a coherent exchange of quantum correlations between the two coupled modes within the waveguides. However, the scenario  changes when we introduce optical loss to the system. It is evident for the figure that increasing the loss rate not only decreases the maximum achievable steering but also reduces the interaction-time interval in which considerable amount steering can be preserved. This is due to the damping term $e^{-\kappa t}$ in the moment equation which are entering the covariance matrix. As photons are lost from the waveguides, the correlations responsible for quantum steering are gradually reduced.
%%%%%%%%%%%%%%%%%%%%%%%%%%%%%%%%%%%%%%%%%%%%%%%%%%%%%%%%%%%%%%%%%%%%%%%%%%%%%%%%%%%%%%%%%%%%

It is worth noting that in Fig.~\ref{fig:4.3}, the oscillatory behavior is still apparent even in the presence of optical dissipation. Thus, optical loss does not immediately destroy the coherent coupling dynamics; rather, it introduces a decaying envelope over the steering oscillations. The competition between $J$ and $\kappa$ is therefore important to the generation of quantum steering dynamics. The coupling parameter $J$ exchanges quantum fluctuations and correlations between the two modes within the waveguides, while the loss parameter $\kappa$ repeatedly eliminates them from the system. Hence, the ratio $\kappa/J$ acts like a measure of the comparative significance of loss and coherent coupling. Smaller values of $\kappa/J$ support the maintenance of quantum steering for longer interaction times, while larger values cause its rapid degradation.

\begin{figure}[t]
	\centering
	\includegraphics[width=2.8in]{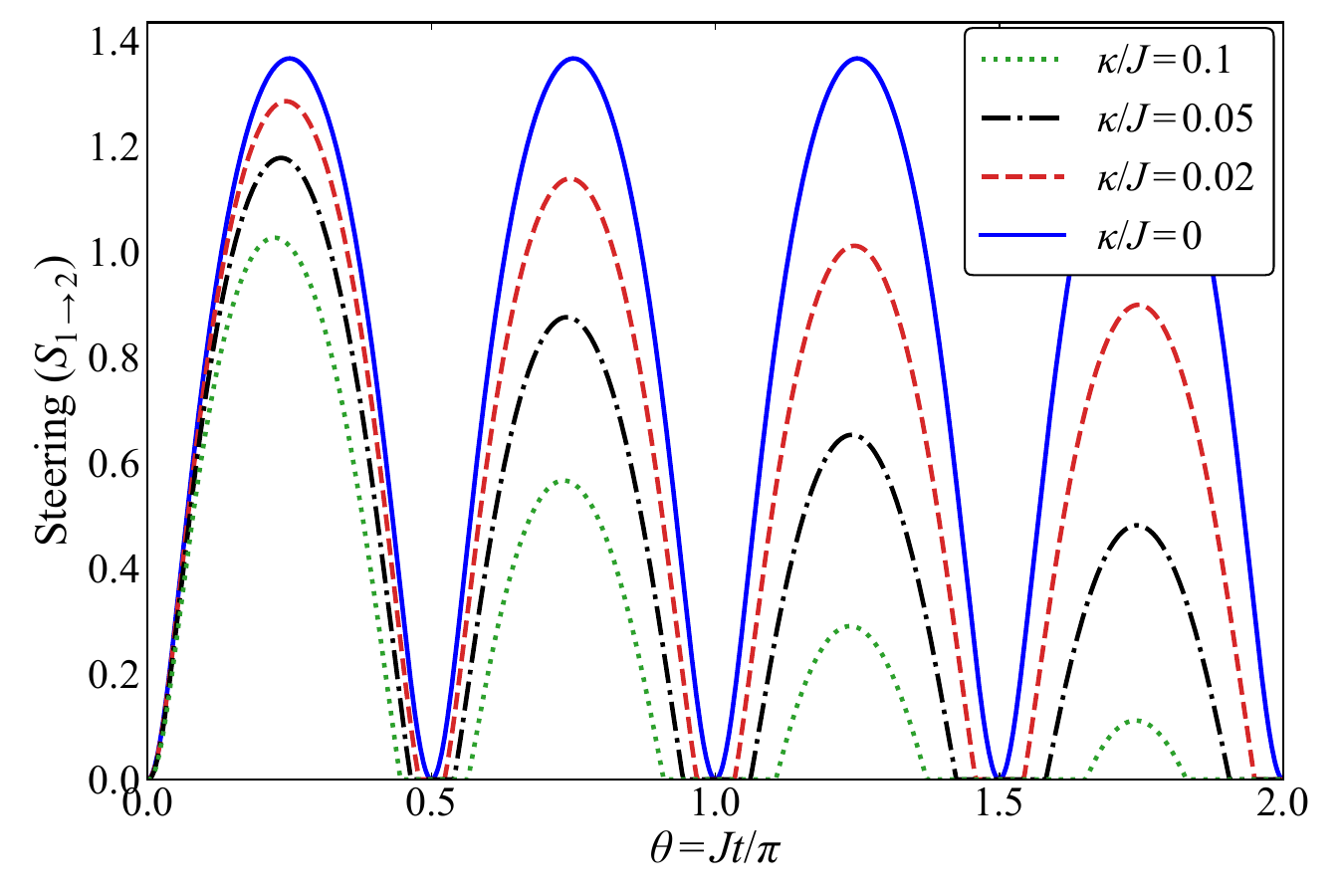}
	\caption{Effect of optical loss with squeezed input states is plotted against dimensionless interaction time, with squeezing factor $r = 0.8$. In each case, the solid blue, dashed-red, dot-dashed black, and dotted-green lines correspond to $\kappa/J = 0, 0.02, 0.05$, and $0.1$, respectively.}
	\label{fig:4.5}
\end{figure}

Next, as an example, we consider squeezed states at the inputs and analyze the influence of optical loss on the generated steering, with the squeezing parameter fixed at $r=0.8$, as shown in Fig.~\ref{fig:4.5}. Not that squeezing provides strong nonclassical quadrature correlations, which are redistributed between the two modes within the waveguides due to the coherent coupling and contribute to the generation of Gaussian steering. From the figure, it is obvious that steering exhibits large and persistent oscillations for $\kappa/J=0$. However, the oscillations gradually decay with the increase of loss, for example, the reduction is relatively small for $\kappa/J=0.02$, becomes more pronounced for $\kappa/J=0.05$, and is strongest for $\kappa/J=0.10$. This behavior demonstrates the competition between squeezing-assisted generation of quantum steering and their degradation by optical loss. It is worth noting that a considerable amount of steering persists over a finite interaction-time interval even for a nonzero value of $\kappa/J$. This indicates that squeezed states at the input offers a degree of robustness against propagation losses. At sufficiently large loss, however, the correlations are progressively suppressed, resulting in a corresponding reduction of Gaussian quantum steering between the two modes.
%%%%%%%%%%%%%%%%%%%%%%%%%%%%%%%%%%%%%%%%%%%%%%%%%%%%%%%%%%%%%%%%%%%%%%%%%%%%%%%%%%%%%%%%%%%%%%%%

\begin{figure}[t]
	\centering
	\includegraphics[width=2.8in]{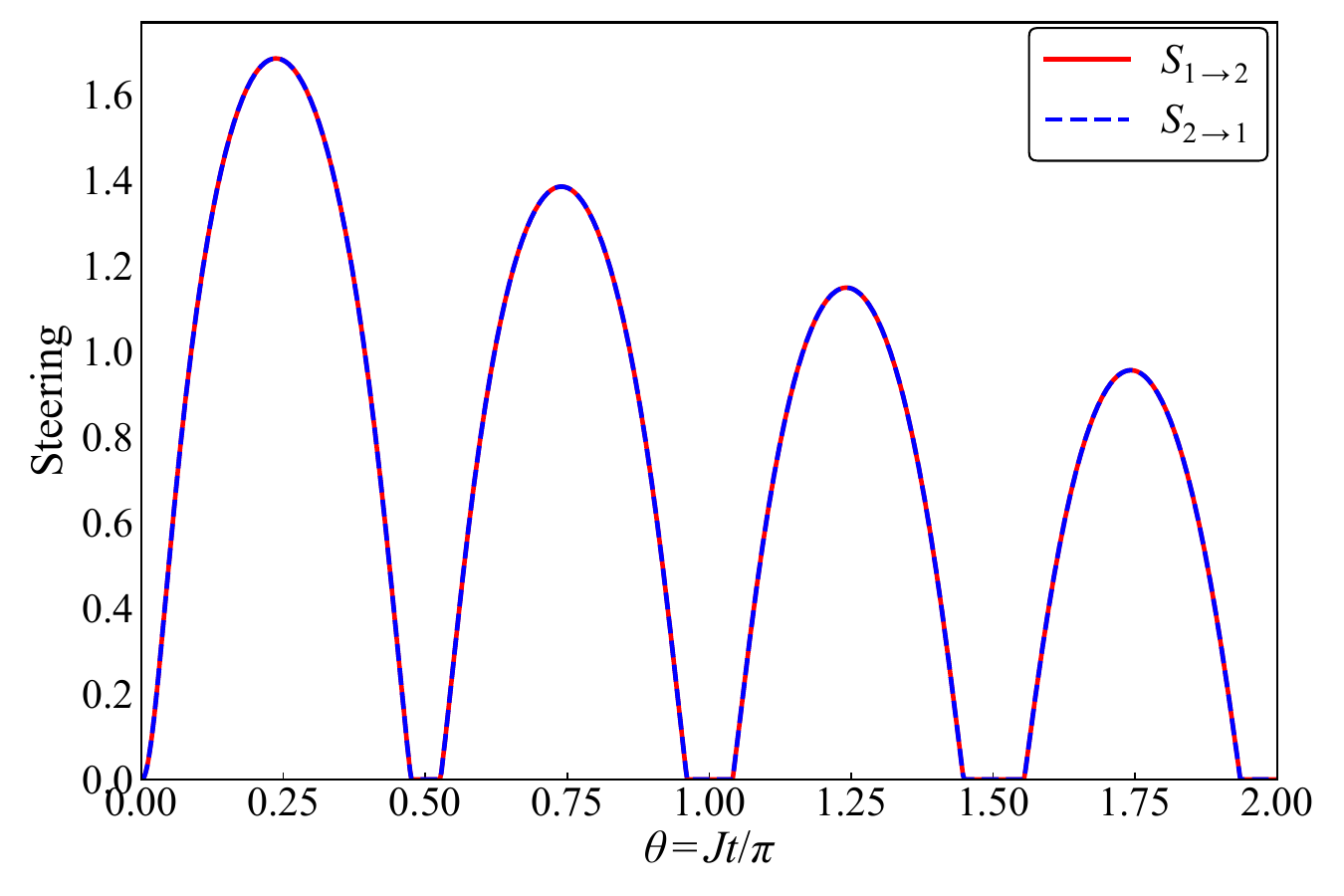}
	\caption{Gaussian quantum steering is plotted against dimensionless interaction time, with equal nonclassicalities of two input modes $\tau_1=\tau_2=0.45$ and $\kappa_1 = \kappa_2 = \kappa = 0.02$, $\mu_1 = \mu_2 = \mu = 1$, $\phi_1 = 0$, and $\phi_2 = \pi/2$.}
	\label{fig:4.7}
\end{figure}

\begin{figure}[t]
	\centering
	%\subfigure{\includegraphics[width=1.73in]{graphs/ASSYMETRY_same tau.pdf}\put(-20,60){\color{black}{$(a)$}}} 
	\subfigure{\includegraphics[width=2.8in]{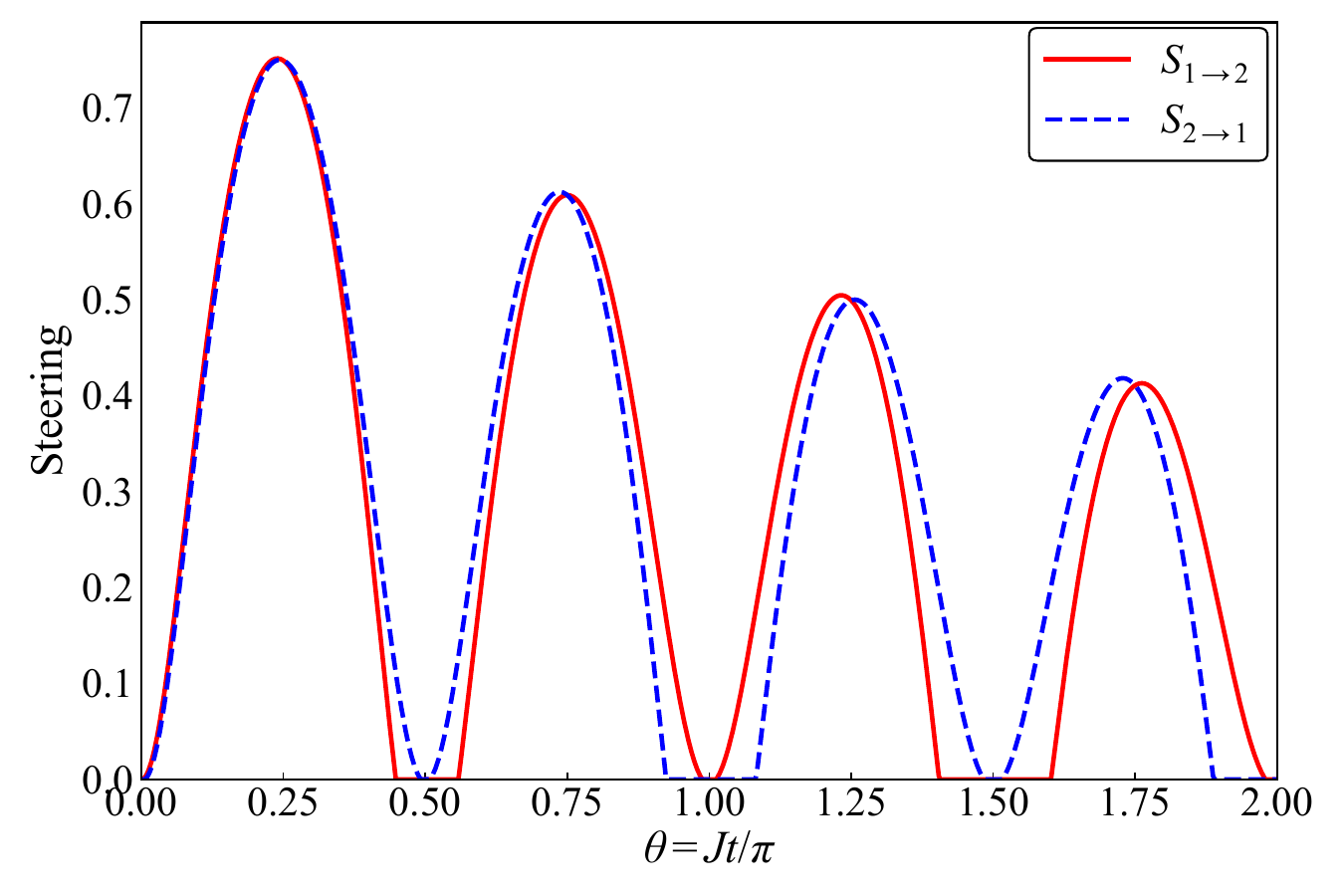}\put(-25,84){\color{black}{$(a)$}}}	
	\subfigure{\includegraphics[width=2.8in]{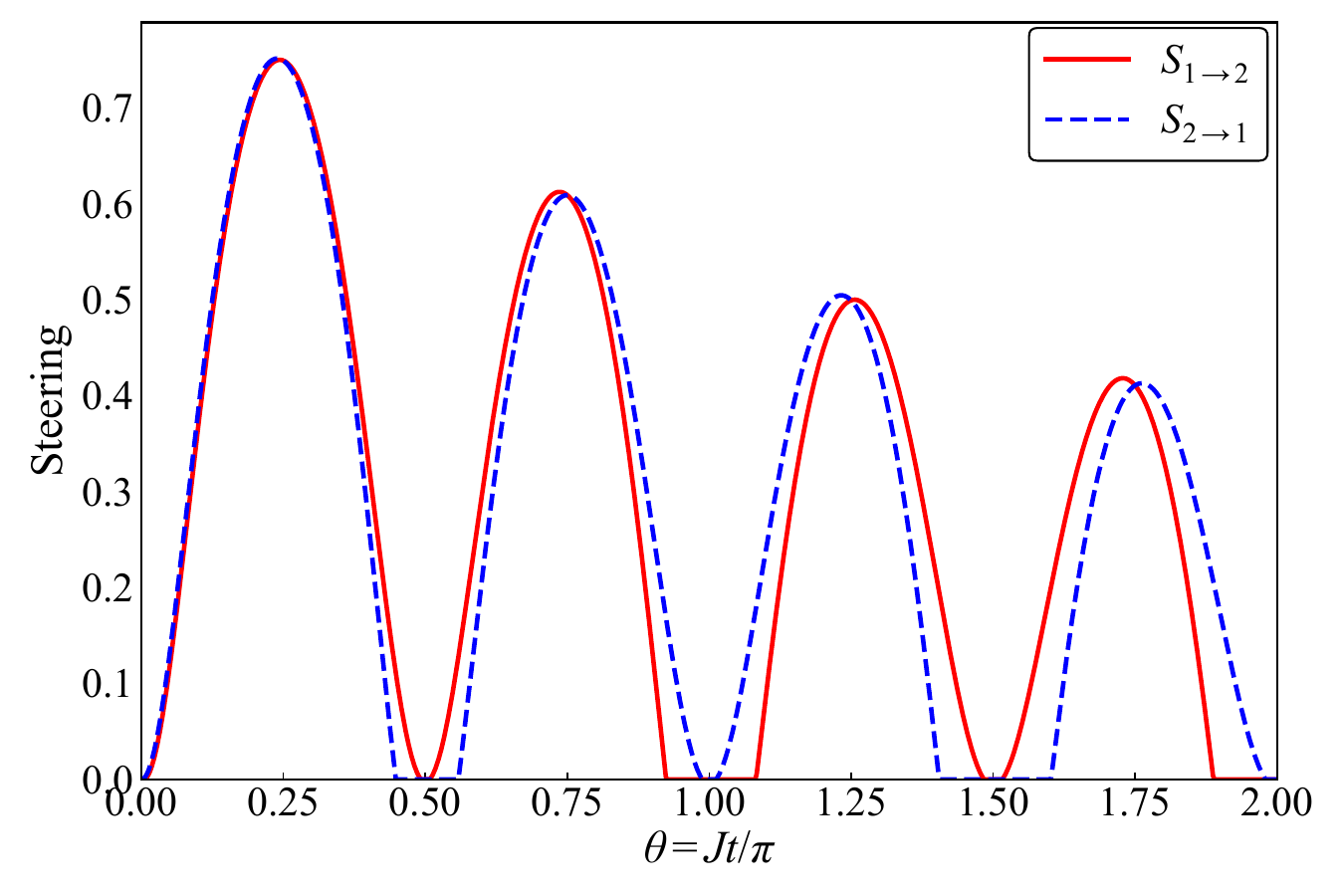}\put(-25,84){\color{black}{$(b)$}}}	
	\caption{Gaussian quantum steering is plotted against dimensionless interaction time, with unequal nonclassicalities of two input modes (a) $\tau_1=0.45$, $\tau_2=0.15$, and (b) $\tau_1=0.15$, $\tau_2=0.45$. Other system parameter values are kept fixed $\kappa_1 = \kappa_2 = \kappa = 0.02$, $\mu_1 = \mu_2 = \mu = 1$, $\phi_1 = 0$, and $\phi_2 = \pi/2$.}
	\label{fig:4.8}
\end{figure}

In addition, we study the directional character of Gaussian quantum steering. It is important to note that steering is intrinsically asymmetric unlike entanglement, and hence the ability of one mode to steer the other can differ from that in the reverse direction. The dynamics of steering between the two modes in the coupled waveguides are analytically analyzed in Sec.~\ref{sec:AsySt}, and graphically illustrated in Figs.~\ref{fig:4.7}--\ref{fig:4.8}. These findings allow us to examine how the preparation of the two input Gaussian modes influence the direction and strength of the generated quantum steering.

In Fig.~\ref{fig:4.7}, we consider equal nonclassicalities and purities of the input Gaussian modes, with identical optical dissipation. As shown analytically in Sec.~\ref{sec:AsySt}, this situation makes $\det M(t)=\det N(t)$, and the steering measures given in Eqs.~\eqref{eq:St(1-2)} and \eqref{eq:St(2-1)} accordingly yield equal strength of steering in the two directions. This determines the symmetric reference case for the directional analysis. Under these conditions, the steering evolution is thus mainly controlled by the coherent coupling parameter $J$. Hence, the steering in the two directions evolves periodically as correlations are exchanged between the modes. This situation offers a useful reference for comparison with the asymmetric input arrangements considered below.
%%%%%%%%%%%%%%%%%%%%%%%%%%%%%%%%%%%%%%%%%%%%%%%%%%%%%%%%%%%%%%%%%%%%%%%%%%

The directional nature of quantum steering becomes more apparent when the nonclassicalities of the two input Gaussian modes are unequal, for instance $\tau_1=0.45$ and $\tau_2=0.15$, as shown Fig.~\ref{fig:4.8}(a). Here, the first mode contains a significantly larger amount of input nonclassicality than the second mode. This imbalance modifies the evolved covariance matrix of the coupled system and, in particular, $\det M(t)\ne \det N(t)$, as analytically explained in Sec.~\ref{sec:AsySt}. Consequently, the measure for two directional steering, given in Eqs.~(\ref{eq:St(1-2)}) and (\ref{eq:St(2-1)}), no longer follow identical dynamics and steerability in the two directions are now different, showing its asymmetric nature. This infers that the redistribution of nonclassical resources between the modes directly influence steering directionality.
%\begin{figure}[t]
%	\centering
%	\includegraphics[width=2.8in]{graphs/ASSYMETRY_HALFKAPPA.pdf}
%	\caption{Gaussian quantum steering is plotted against dimensionless interaction time, with unequal nonclassicalities of two input modes $\tau_1=0.45$, $\tau_2=0.15$, $\kappa_1 = \kappa_2 = \kappa = 0.02$, $\mu_1 = \mu_2 = \mu = 1$, $\phi_1 = 0$, and $\phi_2 = \pi/2$.}
%	\label{fig:4.8}
%\end{figure}

To further establish the connection between input nonclassicality and asymmetry of steering, we consider the reversed configuration, for instance $\tau_1=0.15$ and $\tau_2=0.45$, as presented in Fig.~\ref{fig:4.8}(b). As compared to Fig.~\ref{fig:4.8}(a), the mode containing larger initial nonclassicality is now interchanged. Consequently, the directional nature of steering is accordingly modified. The numerical results hence confirm the analytical origin of the asymmetric two-way steering in the unequal distribution of input nonclassicality (see Sec.~\ref{sec:AsySt}). Figs.~\ref{fig:4.8}(a) and \ref{fig:4.8}(b) hence  demonstrates that steering asymmetry is not simply a fixed property of the coupled waveguide system, but can be controlled through the preparation of the input nonclassical states.
%%%%%%%%%%%%%%%%%%%%%%%%%%%%%%%%%%%%%%%%%%%%%%%%%%%%%%%%

In conclusion, the exchange of correlations due to the coherent coupling may support steering in one or the other direction, or potentially allow steering in both directions. The relative values of $\tau_1$ and $\tau_2$ hence provide a direct control for manipulating the directional nature of Gaussian quantum steering. Such controllability is of special importance in one-way steering~\cite{JBowles14,QZeng22} and asymmetric quantum information protocols~\cite{CBranciard12,NWalk16}, where information or trust is intentionally distributed unequally between two parties. Therefore, the findings in Figs.~\ref{fig:4.7} and \ref{fig:4.8} demonstrate that coupled waveguides provide a flexible and promising platform in which both the strength and direction of Gaussian quantum steering can be controlled through input-state engineering.
%\begin{figure}[t]
%	\centering
%	\includegraphics[width=2.8in]{graphs/ASSYMETRY_tau1_0.15.pdf}
%	\caption{Gaussian quantum steering is plotted against dimensionless interaction time, with unequal nonclassicalities of two input modes $\tau_1=0.15$, $\tau_2=0.45$, $\kappa_1 = \kappa_2 = \kappa = 0.02$, $\mu_1 = \mu_2 = \mu = 1$, $\phi_1 = 0$, and $\phi_2 = \pi/2$.}
%	\label{fig:4.9}
%\end{figure}

\section{Conclusions}\label{conclusion}

In this work, the dynamical behaviour of Gaussian quantum steering has investigated in a coupled lossy waveguide system within the framework of covariance matrix formalism. The main aim of this study was to gain insight into the impact significant physical parameters, for instance, coherent coupling between the waveguide, the nonclassicality and purity of the input states, and the realistic optical dissipation on the persistence and utility of quantum correlations in coupled waveguides system~\cite{Weedbrook2012,Kogias2015,Seifoory2019}. Furthermore, we have explored the intrinsic asymmetric nature of quantum steering when the two input Gaussian modes have unequal nonclassicalities.

The analysis has revealed that quantum steering has oscillatory behaviour with respect to propagation time which is due to the coherent transfer of energy between the coupled waveguide modes. These oscillations are the evidence of the underlying evanescent mechanism of coupling and are a conclusive evidence of interaction between coherent modes within the system. Moreover, on one hand, the increase of the nonclassicality and purity is discovered to increase steering, and thus increase its resistance to optical loss. On the other hand, an increased optical dissipation degraded the quantum steering but still persist for several cycles with respect to the interaction time. The study further indicated that while ideal lossless systems can sustain strong quantum correlations, realistic losses decrease the magnitude of quantum correlations, but steering remains robust even in the presence of optical losses.

In addition, one of the key findings of the present work is the controllable asymmetry of two-way Gaussian quantum steering. We have shown analytically and numerically that, for identical waveguides with equal optical loss, quantum steering from mode 1 to mode 2 and the reverse are not equal specifically when the two input Gaussian modes possess unequal nonclassicality, while identical nonclassicality inputs yield symmetric steering~\cite{WisemanJonesDoherty2007,Uola2020,MHShah26}.

These results are directly relevant to current challenges in quantum information processing, particularly in photonic quantum computing architectures where decoherence and loss severely limit scalability~\cite{Weedbrook2012,Kogias2015,Seifoory2019}. By explicitly modeling dissipation and quantifying the survivability of quantum steering during propagation, this study provides a realistic assessment of how steering behaves under practical conditions. In this sense, the work did not eliminate fundamental limitations such as losses, but instead clarified the parameter regimes in which useful quantum correlations can still be preserved.

In summary, this work has demonstrated that appreciable amounts of Gaussian quantum steering can be maintained in integrated waveguide systems, while its directional character can be coherently controlled.

%\begin{backmatter} 
%	\bmsection{Disclosures}
%	The authors declare no conflicts of interest.
%	\bmsection{Data Availability Statement}
%	Data underlying the results presented in this paper are not publicly available at this time but may be obtained from the authors upon reasonable request.
%\end{backmatter}

\bibliography{St_References}

@article{Harraf2025,
	author  = {Harraf, Hamza and Chabar, Noura and Amazioug, Mohamed and
	Ahl Laamara, Rachid and Haddadi, Saeed},
	title   = {Multipartite {EPR}-steering and entanglement in a cavity
	magnomechanical system through coherent feedback},
	journal = {Sci. Rep.},
	year    = {2025},
	volume  = {15},
	number  = {1},
	pages   = {36742},
	doi = {10.1038/s41598-025-20642-1},
	url = {https://doi.org/10.1038/s41598-025-20642-1},
	issn    = {2045-2322}
}

@article{CBranciard12,
	title = {One-sided device-independent quantum key distribution: Security, feasibility, and the connection with steering},
	author = {Branciard, Cyril and Cavalcanti, Eric G. and Walborn, Stephen P. and Scarani, Valerio and Wiseman, Howard M.},
	journal = {Phys. Rev. A},
	volume = {85},
	issue = {1},
	pages = {010301(R)},
	numpages = {5},
	year = {2012},
	month = {Jan},
	publisher = {American Physical Society},
	doi = {10.1103/PhysRevA.85.010301},
	url = {https://doi.org/10.1103/PhysRevA.85.010301},}

@article{QZeng22,
	title = {Reliable experimental certification of one-way Einstein-Podolsky-Rosen steering},
	author = {Zeng, Qiang and Shang, Jiangwei and Nguyen, H. Chau and Zhang, Xiangdong},
	journal = {Phys. Rev. Res.},
	volume = {4},
	issue = {1},
	pages = {013151},
	numpages = {9},
	year = {2022},
	month = {Feb},
	publisher = {American Physical Society},
	doi = {10.1103/PhysRevResearch.4.013151},
	url = {https://doi.org/10.1103/PhysRevResearch.4.013151},}

@article{NWalk16,
	author = {Nathan Walk and Sara Hosseini and Jiao Geng and Oliver Thearle and Jing Yan Haw and Seiji Armstrong and Syed M. Assad and Jiri Janousek and Timothy C. Ralph and Thomas Symul and Howard M. Wiseman and Ping Koy Lam},
	journal = {Optica},
	number = {6},
	pages = {634--642},
	publisher = {Optica Publishing Group},
	title = {Experimental demonstration of Gaussian protocols for one-sided device-independent quantum key distribution},
	volume = {3},
	month = {Jun},
	year = {2016},
	url = {https://doi.org/10.1364/OPTICA.3.000634},
	doi = {10.1364/OPTICA.3.000634},
}

@article{JBowles14,
	title = {One-way Einstein-Podolsky-Rosen Steering},
	author = {Bowles, Joseph and V\'ertesi, Tam\'as and Quintino, Marco T\'ulio and Brunner, Nicolas},
	journal = {Phys. Rev. Lett.},
	volume = {112},
	issue = {20},
	pages = {200402},
	numpages = {5},
	year = {2014},
	month = {May},
	publisher = {American Physical Society},
	doi = {10.1103/PhysRevLett.112.200402},
	url = {https://doi.org/10.1103/PhysRevLett.112.200402},}

@article{Chae2024,
	author  = {Chae, Eunmi and Choi, Joonhee and Kim, Junki},
	title   = {An elementary review on basic principles and developments of
	qubits for quantum computing},
	journal = {Nano Converg.},
	year    = {2024},
	volume  = {11},
	number  = {1},
	pages   = {11},
	doi = {10.1186/s40580-024-00418-5},
	url = {https://doi.org/10.1186/s40580-024-00418-5},
	issn    = {2196-5404}
}

@article{HSQureshi22_TviaPC,
	doi = {10.1088/1361-6455/ac7370},
	url = {https://doi.org/10.1088/1361-6455/ac7370},
	year = {2022},
	month = {jun},
	publisher = {IOP Publishing},
	volume = {55},
	number = {14},
	pages = {145501},
	author = {Qureshi, Haleema Sadia and Ullah, Shakir and Ghafoor, Fazal},
	title = {Time-dependent quantum teleportation via a parametric converter},
	journal = {J. Phys. B: At. Mol. Opt. Phys.}
}

@article{SchrodingerE35,
  author = {Schrödinger, E.},
  title = {Discussion of probability relations between separated systems},
  journal = {Math. Proc. Camb. Philos. Soc.},
  volume = {31},
  number = {4},
  pages = {555--563},
  url = {https://doi.org/10.1017/S0305004100013554},
  year = {1935},
}

@article{einsteinPodolskyRosen1935,
  author = {Einstein, A. and Podolsky, B. and Rosen, N.},
  title = {Can quantum-mechanical description of physical reality be considered complete?},
  journal = {Phys. Rev.},
  volume = {47},
  number = {10},
  pages = {777--780},
  year = {1935},
  url = {https://doi.org/10.1103/PhysRev.47.777},
}

@article{WisemanJonesDoherty2007,
  author = {Wiseman, H. M. and Jones, S. J. and Doherty, A. C.},
  title = {Steering, entanglement, nonlocality, and the Einstein-Podolsky-Rosen paradox},
  journal = {Phys. Rev. Lett.},
  volume = {98},
  number = {14},
  pages = {140402},
  year = {2007},
  doi = {10.1103/PhysRevLett.98.140402},
  url = {https://doi.org/10.1103/PhysRevLett.98.140402}
}

@article{Weedbrook2012,
  author = {Weedbrook, C. and Pirandola, S. and García-Patrón, R. and Cerf, N. J. and Ralph, T. C. and Shapiro, J. H. and Lloyd, S.},
  title = {Gaussian quantum information},
  journal = {Rev. Mod. Phys.},
  volume = {84},
  number = {2},
  pages = {621},
  year = {2012},
  doi = {10.1103/RevModPhys.84.621},
  url = {https://doi.org/10.1103/RevModPhys.84.621}
}

@article{BraunsteinVanLoock2005,
  author = {Braunstein, S. L. and van Loock, P.},
  title = {Quantum information with continuous variables},
  journal = {Rev. Mod. Phys.},
  volume = {77},
  number = {2},
  pages = {513},
  year = {2005},
  doi = {10.1103/RevModPhys.77.513},
  url = {https://doi.org/10.1103/RevModPhys.77.513}
}

@article{HSQureshiBS18,
	author    = {Qureshi, H. S. and Ullah, S. and Ghafoor, F.},
	title     = {Hierarchy of quantum correlations using a linear beam splitter},
	journal   = {Sci. Rep.},
	volume    = {8},
	number    = {1},
	pages     = {16288},
	year      = {2018},
	doi = {10.1038/s41598-018-34463-y},
	publisher = {Springer Nature},
  url = {https://doi.org/10.1038/s41598-018-34463-y}
}

@article{JKadlec24,
	author = {Josef Kadlec and Karol Bartkiewicz and Anton\'{i}n \v{C}ernoch and Karel Lemr and Adam Miranowicz},
	journal = {Opt. Express},
	number = {2},
	pages = {2333--2346},
	publisher = {Optica Publishing Group},
	title = {Experimental hierarchy of the nonclassicality of single-qubit states via potentials for entanglement, steering, and Bell nonlocality},
	volume = {32},
	month = {Jan},
	year = {2024},
	url = {https://doi.org/10.1364/OE.506169},
	doi = {10.1364/OE.506169},
}

@article{SUllah19,
	author    = {S. Ullah and H. S. Qureshi and F. Ghafoor},
	title     = {Quantum steering of a two-mode Gaussian state using a quantum beat laser},
	journal   = {Appl. Opt.},
	volume    = {58},
	pages     = {7014},
	year      = {2019},
	doi = {10.1364/AO.58.007014},
  url = {https://doi.org/10.1364/AO.58.007014}
}

@article{UllahQureshiGhafoor2019,
  author = {Ullah, S. and Qureshi, H. S. and Ghafoor, F.},
  title = {Hierarchy of temporal quantum correlations using a correlated spontaneous emission laser},
  journal = {Opt. Express},
  volume = {27},
  number = {19},
  pages = {26858--26873},
  year = {2019},
  doi = {10.1364/OE.27.026858},
  url = {https://doi.org/10.1364/OE.27.026858}
}

@article{Kogias2015,
  author = {Kogias, I. and Lee, A. R. and Ragy, S. and Adesso, G.},
  title = {Quantification of Gaussian quantum steering},
  journal = {Phys. Rev. Lett.},
  volume = {114},
  number = {6},
  pages = {060403},
  year = {2015},
  doi = {10.1103/PhysRevLett.114.060403},
  url = {https://doi.org/10.1103/PhysRevLett.114.060403}
}

@article{Politi2008,
  author = {Politi, A. and Cryan, M. J. and Rarity, J. G. and Yu, S. and O'Brien, J. L.},
  title = {Silica-on-silicon waveguide quantum circuits},
  journal = {Science},
  volume = {320},
  number = {5876},
  pages = {646--649},
  year = {2008},
  doi = {10.1126/science.1155441},
  url = {https://doi.org/10.1126/science.1155441}
}

@book{NielsenChuang2000,
  author = {Nielsen, M. A. and Chuang, I. L.},
  title = {Quantum Computation and Quantum Information},
  publisher = {Cambridge University Press},
  address = {Cambridge},
  year = {2000},
  doi = {10.1017/CBO9780511976667},
  url = {https://doi.org/10.1017/CBO9780511976667}
}

@article{OBrien2009,
  author = {O'Brien, J. L. and Furusawa, A. and Vučković, J.},
  title = {Photonic quantum technologies},
  journal = {Nat. Photonics},
  volume = {3},
  number = {12},
  pages = {687--695},
  year = {2009},
  doi = {10.1038/nphoton.2009.229},
  url = {https://doi.org/10.1038/nphoton.2009.229}
}

@article{Peruzzo2010,
  author = {Peruzzo, A. and Lobino, M. and Matthews, J. C. F. and Matsuda, N. and Politi, A. and Poulios, K. and Zhou, X.-Q. and Lahini, Y. and Ismail, N. and W{\"o}rhoff, K. and Bromberg, Y. and Silberberg, Y. and Thompson, M. G. and O'Brien, J. L.},
  title = {Quantum walks of correlated photons},
  journal = {Science},
  volume = {329},
  number = {5998},
  pages = {1500--1503},
  year = {2010},
  doi = {10.1126/science.1193515},
  url = {https://doi.org/10.1126/science.1193515}
}

@book{BreuerPetruccione2002,
  author = {Breuer, H. P. and Petruccione, F.},
  title = {The Theory of Open Quantum Systems},
  publisher = {Oxford University Press},
  year = {2002},
  doi = {10.1093/acprof:oso/9780199213900.001.0001},
  url = {https://doi.org/10.1093/acprof:oso/9780199213900.001.0001}
}

@book{GardinerZoller2004,
  author = {Gardiner, C. W. and Zoller, P.},
  title = {Quantum noise: A handbook of Markovian and non-Markovian quantum stochastic methods with applications to quantum optics},
  year = {2004},
  url = {https://link.springer.com/book/9783540223016}
}

@article{Simon2000,
  author = {Simon, R.},
  title = {Peres-Horodecki separability criterion for continuous variable systems},
  journal = {Phys. Rev. Lett.},
  volume = {84},
  number = {12},
  pages = {2726},
  year = {2000},
  doi = {10.1103/PhysRevLett.84.2726},
  url = {https://doi.org/10.1103/PhysRevLett.84.2726}
}

@article{Uola2020,
	author = {Uola, R. and Costa, A. C. S. and Nguyen, H. C. and G{\"u}hne, O.},
	title = {Quantum steering},
	journal = {Rev. Mod. Phys.},
	volume = {92},
	number = {1},
	pages = {015001},
	year = {2020},
  doi = {10.1103/RevModPhys.92.015001},
  url = {https://doi.org/10.1103/RevModPhys.92.015001}
}

@article{KSun16,
	title = {Experimental Quantification of Asymmetric Einstein-Podolsky-Rosen Steering},
	author = {Sun, Kai and Ye, Xiang-Jun and Xu, Jin-Shi and Xu, Xiao-Ye and Tang, Jian-Shun and Wu, Yu-Chun and Chen, Jing-Ling and Li, Chuan-Feng and Guo, Guang-Can},
	journal = {Phys. Rev. Lett.},
	volume = {116},
	issue = {16},
	pages = {160404},
	numpages = {6},
	year = {2016},
	month = {Apr},
	publisher = {American Physical Society},
	doi = {10.1103/PhysRevLett.116.160404},
	url = {https://doi.org/10.1103/PhysRevLett.116.160404},}

@article{Adesso2007,
  author = {Adesso, G. and Illuminati, F.},
  title = {Entanglement in continuous-variable systems: recent advances and current perspectives},
  journal = {J. Phys. A: Math. Theor.},
  volume = {40},
  number = {28},
  pages = {7821},
  year = {2007},
  doi = {10.1088/1751-8113/40/28/S01},
  url = {https://doi.org/10.1088/1751-8113/40/28/S01}
}

@book{Serafini2017,
  author = {Serafini, A.},
  title = {Quantum Continuous Variables: A Primer of Theoretical Methods},
  publisher = {CRC Press},
  year = {2017},
  doi = {10.1201/9781315118727},
  url = {https://doi.org/10.1201/9781315118727}
}

@article{AdessoRagyLee2014,
  author = {Adesso, G. and Ragy, S. and Lee, A. R.},
  title = {Continuous variable quantum information: Gaussian states and beyond},
  journal = {Open Syst. Inf. Dyn.},
  volume = {21},
  number = {1},
  pages = {1440001},
  year = {2014},
  url = {https://doi.org/10.1142/S1230161214400010}
}

@article{Yariv1973,
  author = {Yariv, A.},
  title = {Coupled-mode theory for guided-wave optics},
  journal = {IEEE J. Quantum Electron.},
  volume = {9},
  number = {9},
  pages = {919--933},
  year = {1973},
  doi = {10.1109/JQE.1973.1077767},
  url = {https://doi.org/10.1109/JQE.1973.1077767}
}

@article{PianiWatrous2015,
  author = {Piani, M. and Watrous, J.},
  title = {Necessary and sufficient quantum information characterization of Einstein-Podolsky-Rosen steering},
  journal = {Phys. Rev. Lett.},
  volume = {114},
  number = {6},
  pages = {060404},
  year = {2015},
  doi = {10.1103/PhysRevLett.114.060404},
  url = {https://doi.org/10.1103/PhysRevLett.114.060404}
}

@article{Crespi2013,
  author = {Crespi, A. and Ramponi, R. and Osellame, R. and Sansoni, L. and Bonelli, I. and Sciarrino, F. and Vallone, G. and Mataloni, P.},
  title = {Integrated photonic quantum gates for polarization qubits},
  journal = {Nat. Commun.},
  volume = {2},
  pages = {566},
  year = {2011},
  doi = {10.1038/ncomms1570},
  url = {https://doi.org/10.1038/ncomms1570}
}

@article{Ekert1991,
  author = {Ekert, A. K.},
  title = {Quantum cryptography based on Bell's theorem},
  journal = {Phys. Rev. Lett.},
  volume = {67},
  pages = {661--663},
  year = {1991},
  doi = {10.1103/PhysRevLett.67.661},
  url = {https://doi.org/10.1103/PhysRevLett.67.661}
}

@article{BennettBrassardCrepeauJozsaPeresWootters1993,
  author = {Bennett, C. H. and Brassard, G. and Cr{\'e}peau, C. and Jozsa, R. and Peres, A. and Wootters, W. K.},
  title = {Teleporting an unknown quantum state via dual classical and Einstein--Podolsky--Rosen channels},
  journal = {Phys. Rev. Lett.},
  volume = {70},
  pages = {1895--1899},
  year = {1993},
  doi = {10.1103/PhysRevLett.70.1895},
  url = {https://doi.org/10.1103/PhysRevLett.70.1895}
}

@book{SalehTeich2007,
  author = {Saleh, B. E. A. and Teich, M. C.},
  title = {Fundamentals of Photonics},
  edition = {2nd},
  publisher = {Wiley-Interscience},
  address = {Hoboken, NJ, USA},
  year = {2007},
  url = {https://katalog.bibliothek.kit.edu/bib/1062330}
}

@article{Wenger2004,
	author  = {Wenger, J. and Fiur{\'a}{\v{s}}ek, J. and Tualle-Brouri, R. and Cerf, N. J. and Grangier, P.},
	title   = {Pulsed squeezed vacuum characterization without homodyning},
	journal = {Phys. Rev. A},
	volume  = {70},
	pages   = {053812},
	year    = {2004},
	doi = {10.1103/PhysRevA.70.053812},
	url = {https://doi.org/10.1103/PhysRevA.70.053812},}

@article{Dauria2009,
	author  = {D'Auria, V. and Fornaro, S. and Porzio, A. and Solimeno, S. and Olivares, S. and Paris, M. G. A.},
	title   = {Full Characterization of Gaussian Bipartite Entangled States by a Single Homodyne Detector},
	journal = {Phys. Rev. Lett.},
	volume  = {102},
	pages   = {020502},
	year    = {2009},
	doi = {10.1103/PhysRevLett.102.020502},
	url = {https://doi.org/10.1103/PhysRevLett.102.020502},}

@article{Bromberg2009,
  author = {Bromberg, Y. and Lahini, Y. and Morandotti, R. and Silberberg, Y.},
  title = {Quantum and classical correlations in waveguide lattices},
  journal = {Phys. Rev. Lett.},
  volume = {102},
  pages = {253904},
  year = {2009},
  doi = {10.1103/PhysRevLett.102.253904},
  url = {https://doi.org/10.1103/PhysRevLett.102.253904}
}

@article{Rai2010,
  author = {Rai, A. and Das, S. and Agarwal, G. S.},
  title = {Quantum entanglement in coupled lossy waveguides},
  journal = {Opt. Express},
  volume = {18},
  number = {6},
  pages = {6241--6254},
  year = {2010},
  doi = {10.1364/OE.18.006241},
  url = {https://doi.org/10.1364/OE.18.006241}
}

@article{Buono2010,
	author  = {Buono, D. and Nocerino, G. and D'Auria, V. and Porzio, A. and Olivares, S. and Paris, M. G. A.},
	title   = {Quantum characterization of bipartite Gaussian states},
	journal = {J. Opt. Soc. Am. B},
	volume  = {27},
	number  = {6},
	pages   = {A110--A118},
	year    = {2010},
	doi = {10.1364/JOSAB.27.00A110},
	url = {https://doi.org/10.1364/JOSAB.27.00A110},}

@article{Seifoory2019,
  author = {Seifoory, H. and Helt, L. G. and Sipe, J. E. and Dignam, M. M.},
  title = {Counterpropagating continuous-variable entangled states in lossy coupled-cavity optical waveguides},
  journal = {Phys. Rev. A},
  volume = {100},
  pages = {033839},
  year = {2019},
  doi = {10.1103/PhysRevA.100.033839},
  url = {https://doi.org/10.1103/PhysRevA.100.033839}
}

@article{Li2006,
  author = {Li, H.-R. and Li, F.-L. and Yang, Y.},
  title = {Entangling two single-mode Gaussian states by use of a beam splitter},
  journal = {Chin. Phys.},
  volume = {15},
  number = {12},
  pages = {2947--2952},
  year = {2006},
  doi = {10.1088/1009-1963/15/12/030},
  url = {https://doi.org/10.1088/1009-1963/15/12/030}
}

@article{HSQureshi23,
	doi = {10.1088/1402-4896/acc3cc},
	url = {https://doi.org/10.1088/1402-4896/acc3cc},
	year = {2023},
	month = {mar},
	publisher = {IOP Publishing},
	volume = {98},
	number = {4},
	pages = {045113},
	author = {Qureshi, Haleema Sadia and Ullah, Shakir and Ghafoor, Fazal},
	title = {Coherence controlled generation of Gaussian quantum discord in a quantum beat laser},
	journal = {Phys. Scr.}
}

@article{NChabarSUllah26,
	author  = {Chabar, Noura and Amghar, M'bark and Ullah, Shakir and Amazioug, Mohamed and Nisar, Kottakkaran Sooppy and Zakarya, Mohammed and Ismail, Gamal M. and Abdel-Aty, Abdel-Haleem},
	title   = {Barnett effect-induced nonreciprocal entanglement and {Gaussian} interferometric power in magnomechanics with optical parametric amplifier},
	journal = {Quantum Inf. Process.},
	year    = {2026},
	volume  = {25},
	number  = {8},
	pages   = {247},
	doi = {10.1007/s11128-026-05252-8},
	url = {https://doi.org/10.1007/s11128-026-05252-8},
	issn    = {1573-1332}
}

@article{HSQureshi21_ENinCEL,
	title={Generation and Control of Bipartite Entanglement in a Correlated Spontaneous-Emission Laser},
	author={Qureshi, Haleema Sadia and Ullah, Shakir and Ghafoor, Fazal},
	journal={J. Russ. Laser Res.},
	volume={42},
	number={5},
	pages={501--511},
	year={2021},
	publisher={Springer},
	url = {https://doi.org/10.1007/s10946-021-09988-9},
  doi = {10.1007/s10946-021-09988-9}
}

@article{MAmazioug18,
	title={Entanglement, EPR steering and Gaussian geometric discord in a double cavity optomechanical systems},
	author={Amazioug, Mohamed and Nassik, Mostafa and Habiballah, Nabil},
	journal={Eur. Phys. J. D.},
	volume={72},
	pages={1--9},
	year={2018},
	publisher={Springer},
	url = {https://doi.org/10.1140/epjd/e2018-90151-6},
  doi = {10.1140/epjd/e2018-90151-6}
}

@article{Iwanow2004,
	author  = {Iwanow, R. and Schiek, R. and Stegeman, G. I. and Pertsch, T. and Lederer, F. and Min, Y. and Sohler, W.},
	title   = {Observation of Discrete Quadratic Solitons},
	journal = {Phys. Rev. Lett.},
	volume  = {93},
	number  = {11},
	pages   = {113902},
	year    = {2004},
	doi = {10.1103/PhysRevLett.93.113902},
	url = {https://doi.org/10.1103/PhysRevLett.93.113902},}

@article{Peschel2002,
	author  = {Peschel, U. and Morandotti, R. and Arnold, J. M. and Aitchison, J. S. and Eisenberg, H. S. and Silberberg, Y. and Pertsch, T. and Lederer, F.},
	title   = {Optical Discrete Solitons in Waveguide Arrays. 2. Dynamic Properties},
	journal = {J. Opt. Soc. Am. B},
	volume  = {19},
	number  = {11},
	pages   = {2637--2644},
	year    = {2002},
	doi = {10.1364/JOSAB.19.002637},
	url = {https://doi.org/10.1364/JOSAB.19.002637},}

@article{Mogensen2004,
	author  = {Mogensen, K. B. and Eriksson, F. and Gustafsson, O. and Nikolajsen, R. P. H. and Kutter, J. P.},
	title   = {Pure-silica optical waveguides, fiber couplers, and high-aspect ratio submicrometer channels for electrokinetic separation devices},
	journal = {Electrophoresis},
	volume  = {25},
	number  = {21--22},
	pages   = {3788--3795},
	year    = {2004},
	doi = {10.1002/elps.200406077},
	url = {https://doi.org/10.1002/elps.200406077},}

@article{HSQureshi20,
	author = {Haleema Sadia Qureshi and Shakir Ullah and Fazal Ghafoor},
	journal = {Appl. Opt.},
	number = {9},
	pages = {2701--2708},
	publisher = {Optica Publishing Group},
	title = {Bipartite Gaussian quantum steering, entanglement, and discord and their interconnection via a parametric down-converter},
	volume = {59},
	month = {Mar},
	year = {2020},
	url = {https://doi.org/10.1364/AO.378891},
	doi = {10.1364/AO.378891},
}

@article{MHShah26,
	author = {Mahboob Hasan Shah and Haleema Sadia Qureshi and Shakir Ullah and Fazal Ghafoor},
	journal = {J. Opt. Soc. Am. B},
	number = {8},
	pages = {B100--B109},
	publisher = {Optica Publishing Group},
	title = {Continuous-variable quantum steering in a Raman-driven quantum beat laser},
	volume = {43},
	month = {Aug},
	year = {2026},
	url = {https://doi.org/10.1364/JOSAB.596470},
	doi = {10.1364/JOSAB.596470},
}

@article{SUllah19_EninRDQBL,
	author    = {Ullah, S. and Qureshi, H. S. and Tiaz, G. and Ghafoor, F. and Saif, F.},
	title     = {Coherence control of entanglement dynamics of two-mode Gaussian state via Raman driven quantum beat laser using Simon’s criterion},
	journal   = {Appl. Opt.},
	volume    = {58},
	pages     = {197},
	year      = {2019},
	doi = {10.1364/AO.58.000197},
  url = {https://doi.org/10.1364/AO.58.000197}
}

@article{NChabarHSQureshi26,
	title = {Nonreciprocal entanglement, quantum synchronization, and optimal fidelity of teleportation in hybrid cavity-magnon optomechanics via the Barnett effect},
	journal = {Chaos Solitons Fractals},
	volume = {209},
	pages = {118560},
	year = {2026},
	issn = {0960-0779},
	doi = {10.1016/j.chaos.2026.118560},
	url = {https://doi.org/10.1016/j.chaos.2026.118560},
	author = {Noura Chabar and Mohamed Amazioug and Haleema Sadia Qureshi and Nazek Alessa and Abdel-Haleem Abdel-Aty}
}

\end{document}